\documentclass[conference]{IEEEtran}
\usepackage{amsfonts}
\usepackage{cite}

\ifCLASSINFOpdf
\else
\usepackage[dvips]{graphicx}
\fi
\usepackage[cmex10]{amsmath}

\begin{document}
%
\title{Carrier Frequency Offset Estimation Approach for Multicarrier Transmission on Hexagonal Time-Frequency Lattice}
%
%
%

\author{\IEEEauthorblockN{Kui~Xu,~Wenfeng~Ma,~Lianguo~Wu,~Wei~Xie,~Dongmei~Zhang,~Youyun~Xu}
\IEEEauthorblockA{Institute of Communications Engineering, PLAUST\\
E-mail:lgdxxukui@126.com}}
\markboth{Journal of \LaTeX\ Class Files,~Vol.~6, No.~1, June~2012}%
{Shell \MakeLowercase{\textit{et al.}}: Bare Demo of IEEEtran.cls
for Journals}
%



\maketitle

\begin{abstract}
In this paper, a novel carrier frequency offset estimation approach,
including preamble structure, carrier frequency offset estimation
algorithm, is proposed for hexagonal multi-carrier transmission
(HMCT) system. The closed-form Cramer-Rao lower bound of the
proposed carrier frequency offset estimation scheme is given.
Theoretical analyses and simulation results show that the proposed
preamble structure and carrier frequency offset estimation algorithm
for HMCT system obtains an approximation to the Cramer-Rao lower
bound mean square error (MSE) performance over the doubly dispersive
(DD) propagation channel.
\end{abstract}

\begin{IEEEkeywords}
Hexagonal Multi-Carrier Transmission System; Preamble Structure;
Carrier Frequency Offset Estimation; Cramer-Rao Lower Bound;
\end{IEEEkeywords}

%
\IEEEpeerreviewmaketitle

\section{Introduction}
\IEEEPARstart{O}{rthogonal} frequency division multiplexing (OFDM)
systems with guard-time interval or cyclic prefix can prevent
inter-symbol interference (ISI). OFDM has overlapping spectra and
rectangular impulse responses. Consequently, each OFDM sub-channel
exhibits a sinc-shape frequency response. Therefore, the time
variations of the channel during one OFDM symbol duration destroy
the orthogonality of different subcarriers, and result in power
leakage among subcarriers, known as inter-carrier interference
(ICI), which causes degradation in system performance. In order to
overcome the above drawbacks of OFDM system, several pulse-shaping
OFDM systems were proposed
\cite{Kumb07,Das07,Abb10,Gao11,Jun07,Lin08,Sio02,Ma08}.

It is shown that signal transmission through a rectangular lattice
is suboptimal for doubly dispersive (DD) channel
\cite{Str03,Yua08,Han07,Han09,Han10}. By using results from sphere
covering theory, the authors have demonstrated that hexagonal
multi-carrier transmission (HMCT) system obtains lower energy
perturbation than OFDM system, hence providing better performance
against ISI/ICI \cite{Str03,Yua08,Han07}. There are mainly two types
of HMCT systems, (1): HMCT system with orthogonal prototype
pulse\cite{Str03, Yua08, Ma08}, named as lattice OFDM (LOFDM)
system, which confines the transmission pulses to a set of
orthogonal ones. (2): HMCT system with nonorthogonal prototype
pulse\cite{Han07,Han09,Han10}, named as hexagonal multicarrier
transmission (HMT) system, which abandons the orthogonality
condition of the modulated pulses and obtains the high robustness
performance on combating the dispersion of DD propagation channel in
both time and frequency domain.

To optimally combat the impact of the DD propagation channels, the
lattice parameters in the time-frequency (T-F) plane and the pulse
shape of modulation waveform of HMCT system are jointly optimized to
adapt to the channel scattering function. It is shown in
\cite{Xu09,Xu11,Xu12,Xu12-1} that the HMCT system outperforms OFDM
system from the robustness against channel dispersion point of view.

The basic mathematical operation of the received signal performed by
the demodulator is a projection onto an identically structured
function set generated by the prototype pulse function \cite{Jun07},
i.e. an optimal match filter. In HMCT system, there is no cyclic
prefix and data symbols of HMCT signal are transmitted on hexagonal
lattice points in the T-F plane. Traditional timing and frequency
synchronization schemes can not be applied to HMCT system directly.
In our previous work \cite{Xu09,Xu11,Xu12,Xu12-1}, signal to
interference and noise ratio (SINR) of HMCT system over DD channel
is analyzed and the Max-SINR based timing synchronization scheme is
proposed.

In this paper, a novel preamble structure and two carrier frequency
offset estimation algorithms, named as phase differential algorithm
and least squares algorithm, for HMCT system are proposed. After
detailed derivation, the close form performance lower bound of the
proposed carrier frequency offset estimation algorithm is given.
Theoretical analyses and simulation results show that the proposed
carrier frequency estimation approach for HMCT system obtains an
approximation to the lower bound mean square error (MSE) performance
over DD channel.

\section{Hexagonal Multi-Carrier Transmission System}
In HMCT systems, the transmitted baseband signal can be expressed as
\cite{Han07,Han09,Han10}
\begin{equation} \label{1}
\begin{split}
x(t)&=\sum_{m}\sum_{n=0}^{\frac{N}{2}-1}c_{m,2n}\psi_{m,2n}(t)+\sum_{m}\sum_{n=0}^{\frac{N}{2}-1}c_{m,2n+1}\psi_{m,2n+1}(t)
\end{split}
\end{equation}
where $T$ and $F$ are the lattice parameters, which can be viewed as
the symbol period and the subcarrier separation, respectively;
$c_{m,n}$ denotes the transmitted data symbol, which is assumed to
be taken from a specific signal constellation and independent and
identically distributed (i.i.d.) with zero mean and average power
$\sigma_{c}^{2}$; $m\in\mathcal {M}$ and $n\in\mathcal {N}$ are the
position indices in the T-F plane; $\mathcal {M}$ and $\mathcal {N}$
denote the sets from which $m,n$ can be taken, with cardinalities
$\textit{M}$ and $\textit{N}$, respectively.
$\psi_{m,2n}(t)=\psi(t-mT)e^{j2\pi nFt}$ and
$\psi_{m,2n+1}(t)=\psi(t-mT-\frac{T}{2})e^{j2\pi(nF+\frac{F}{2})t}$
are the transmitted pulses generated by the prototype pulse
$\psi(t)$. The ambiguity function of the prototype pulse is defined
as
\begin{equation} \label{2}
A_{\psi}(\tau,\upsilon)=\int_{-\infty}^{\infty}\psi(t)\psi^{*}(t-\tau)e^{-j2\pi\upsilon
t}dt
\end{equation}
The baseband DD channel can be modeled as a random linear operator
$\textrm{H}$ \cite{Bel63}
\begin{equation} \label{3}
\textrm{H}[x(t)]=\int_{0}^{\tau_{\textrm{max}}}\int^{f_{d}}_{-f_{d}}H(\tau,\upsilon)x(t-\tau)e^{j2\pi\upsilon
t}d\tau d\upsilon
\end{equation}
where $\tau_{\textrm{max}}$ and $f_{d}$ are the maximum multipath
delay spread and the maximum Doppler frequency,
respectively\cite{Coh95}. $H(\tau,\upsilon)$ is called the
delay-Doppler spread function, which is the Fourier transform of the
time-varying impulse response of the channel $h(t,\tau)$ with
respect to $t$\cite{Bel63}.

In wide-sense stationary uncorrelated scattering (WSSUS) assumption,
the DD channel is characterized by the second-order statistics
\begin{equation} \label{4}
\textmd{E}[H(\tau,\upsilon)H^{*}(\tau_{1},\upsilon_{1})]=S_{H}(\tau,\upsilon)\delta(\tau-\tau_{1})\delta(\upsilon-\upsilon_{1})
\end{equation}
where $\textmd{E}[\cdot]$ denotes the expectation and
$S_{H}(\tau,\upsilon)$ is called the scattering function, which
characterizes the statistics of the WSSUS channel. Without loss of
generality, $H(\tau,\upsilon)$ is assumed to have zero mean and unit
variance,
i.e.,$\int_{0}^{\tau_{\textrm{max}}}\int_{-f_{d}}^{f_{d}}S_{H}(\tau,\upsilon)d\tau
d\upsilon=1$. The received baseband signal can be expressed as
\begin{equation} \label{5}
r(t)=\textrm{H}[x(t)]+w(t)
\end{equation}
where $w(t)$ is the AWGN with variance $\sigma_{w}^{2}$.

The basic mathematical operation of the received signal performed by
the demodulator is a projection onto an identically structured
function set generated by the prototype pulse function, i.e. an
optimal match filter \cite{Xu12-1}. To recover the transmitted data
symbol ${\hat{c}}_{m,2n}$, the match filter receiver projects the
received signal $r(t)$ on prototype pulse function $\psi_{m,2n}(t)$,
i.e., $\hat{c}_{m,2n}=\langle
r(t),\psi_{m,2n}(t)\rangle=\int_{-\infty}^{\infty}r(t)\psi_{m,2n}^{\ast}(t)dt$,
where $\langle\cdot\rangle$ denotes the inner product and
$(\cdot)^{*}$ denotes the complex conjugate. Hence, frequency
synchronization plays a critical role in ensuring reliable
demodulation. Firstly, we will analyze the effects of carrier
frequency offset on the recovered data symbol $\hat{c}_{m,2n}$.

\section{Effects of Carrier Frequency Offset on HMCT System}
Under the assumption that there is a carrier frequency offset
$\Delta f$ between the received signal $r(t)$ and the transmitted
signal $x(t)$. After ignoring the impact of additive noise, the
received signal $r(t)$ can be written as
\begin{equation} \label{6}
\begin{split}
r(t)&=e^{j2\pi\Delta
ft}\int_{0}^{\tau_{\textrm{max}}}\int^{f_{d}}_{-f_{d}}H(\tau,\upsilon)x(t-\tau)e^{j2\pi\upsilon
t}d\tau d\upsilon
\end{split}
\end{equation}
and the recovered data symbol ${\hat{c}}_{m,2n}$ can be expressed as
\begin{equation} \label{7}
\begin{split}
\hat{c}_{m,2n}&=\big<r(t),\psi_{m,2n}(t)\big> \\
 &=\sum_{m'}\sum_{n'=0}^{N/2-1}c_{m',2n'}\Xi_{m,n;m',2n'}^{\Delta
 f}\\
 &+\sum_{m'}\sum_{n'=0}^{N/2-1}c_{m',2n'+1}\Xi_{m,n;m',2n'+1}^{\Delta
 f}
\end{split}
\end{equation}
where $\Xi_{m,n;m',2n'}^{\Delta
 f}$ in (7) can be expressed as (8) at the top of the next page.

\newcounter{mytempeqncnt}
\begin{figure*}
\normalsize \setcounter{mytempeqncnt}{\value{equation}}
\setcounter{equation}{7}
\begin{equation} \label{8}
\begin{split}
\Xi_{m,2n;m',2n'}^{\Delta
 f}&=\int_{-\infty}^{\infty}\int_{0}^{\tau_{\textrm{max}}}\int^{f_{d}}_{-f_{d}}e^{j2\pi\Delta
ft}\psi_{m',2n'}(t-\tau)H(\tau,\upsilon)\psi_{m,2n}^{\ast}(t)e^{j2\pi\upsilon
t}d\tau d\upsilon dt\\
&=e^{-j2\pi FTm(n-n')}e^{-j2\pi mT\Delta
f}\int_{0}^{\tau_{\textrm{max}}}\int^{f_{d}}_{-f_{d}}A_{\psi}^{\ast}\big((m'-m)T+\tau,
(n'-n)F+\upsilon+\Delta f\big)\\
&\cdot H(\tau,\upsilon)e^{-j2\pi n'F\tau}e^{j2\pi \upsilon mT}d\tau
d\upsilon
\end{split}
\end{equation}
\setcounter{equation}{\value{mytempeqncnt}} \hrule
\end{figure*}

Let $m=m'$ and $n=n'$, $\Xi_{m,2n;m,2n}^{\Delta
 f}$ can be expressed
\setcounter{equation}{8}
\begin{equation} \label{9}
\begin{split}
\Xi_{m,2n;m,2n}^{\Delta
 f}&=e^{-j2\pi mT\Delta
f}\int_{0}^{\tau_{\textrm{max}}}\int^{f_{d}}_{-f_{d}}A_{\psi}^{\ast}\big(\tau,
\upsilon+\Delta f\big)\\
&\cdot H(\tau,\upsilon)e^{-j2\pi nF\tau}e^{j2\pi \upsilon mT}d\tau
d\upsilon\\
&=e^{-j2\pi mT\Delta f}A_{H}(\tau_{\textrm{max}},f_d,\Delta f)
\end{split}
\end{equation}
and $\Xi_{m,n;m',2n'+1}^{\Delta
 f}$ in (7) can be expressed as (10) at the top of the next page.

\begin{figure*}
\normalsize \setcounter{mytempeqncnt}{\value{equation}}
\setcounter{equation}{9}
\begin{equation} \label{10}
\begin{split}
\Xi_{m,2n;m',2n'+1}^{\Delta
 f}&=\int_{-\infty}^{\infty}\int_{0}^{\tau_{\textrm{max}}}\int^{f_{d}}_{-f_{d}}e^{j2\pi\Delta
ft}\psi_{m',2n'+1}(t-\tau)H(\tau,\upsilon)\psi_{m,2n}^{\ast}(t)e^{j2\pi\upsilon
t}d\tau d\upsilon dt\\
&=e^{-j2\pi FTm(n-n')}e^{-j2\pi mT\Delta
f}\int_{0}^{\tau_{\textrm{max}}}\int^{f_{d}}_{-f_{d}}A_{\psi}^{\ast}\big((m'-m)T+\tau,
(n'+\frac{1}{2}-n)F+\upsilon+\Delta f\big)\\
& \cdot H(\tau,\upsilon)e^{-j2\pi (n'+\frac{1}{2})F\tau}e^{j2\pi
\upsilon mT}d\tau d\upsilon
\end{split}
\end{equation}
\setcounter{equation}{\value{mytempeqncnt}} \hrule
\end{figure*}

Let $m=m'$ and $n=n'$, $\Xi_{m,2n;m,2n+1}^{\Delta
 f}$ can be expressed as
\setcounter{equation}{10}
\begin{equation} \label{11}
\begin{split}
\Xi_{m,2n;m,2n+1}^{\Delta
 f}&=e^{j\pi mFT}e^{-j2\pi mT\Delta
f}\int_{0}^{\tau_{\textrm{max}}}\int^{f_{d}}_{-f_{d}}\\
&\cdot A_{\psi}^{\ast}\big(\tau, \upsilon+\Delta
f\big)H(\tau,\upsilon)e^{-j2\pi (n+\frac{1}{2})F\tau}\\
&\cdot e^{j2\pi \upsilon (m+\frac{1}{2})T}d\tau d\upsilon
\end{split}
\end{equation}
Hence, the data symbol $\hat{c}_{m,2n}$ in (7) can be rewritten as
\begin{equation} \label{12}
\begin{split}
\hat{c}_{m,2n}&=c_{m,2n}\Xi_{m,2n;m,2n}^{\Delta
 f}+\sum_{n'=0,n'\neq n}^{N/2-1}c_{m,2n'}\Xi_{m,2n;m,2n'}^{\Delta
 f}\\
&+\sum_{m'\neq
0}\sum_{n'=0}^{N/2-1}c_{m',2n'}\Xi_{m,2n;m',2n'}^{\Delta
 f}\\
 &+\sum_{m'}\sum_{n'=0}^{N/2-1}c_{m',2n'+1}\Xi_{m,2n;m',2n'+1}^{\Delta
 f}\\
\end{split}
\end{equation}
The first term in equation (12) represents the desired symbol and
the last three terms denote ISI/ICI. Concerning the useful portion,
the transmitted symbols $c_{m,2n}$ are attenuated by
$A_{H}(\tau_{\textrm{max}},f_d,\Delta f)$ which is caused by the
carrier frequency offset $\Delta f$ and doubly dispersive channel.
Meanwhile, the transmitted symbols rotated by a phasor $-j2\pi
mT\Delta f$.

\section{The Proposed Preamble Structure}
The proposed preamble is composed of two training sequences
$\textit{\textbf{P}}_1$ and $\textit{\textbf{P}}_2$ in the frequency
domain, as depicted in Fig. 1.
$\textit{\textbf{P}}_i=[P_i(0),P_i(1),\cdots,P_i(N_{P}-1)]$, $i \in
\{1,2\}$ and $N_{P}\leq N/2$ denotes the length of training
sequence.

\begin{figure}[!t] \centering
\includegraphics[width=2.8in]{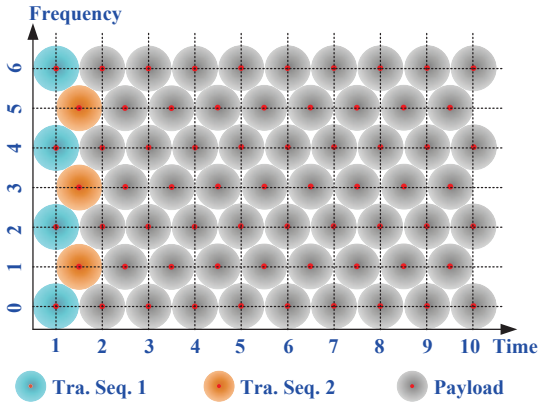}
\caption{The proposed preamble structure.}
\end{figure}

In this paper, $\textit{\textbf{P}}_i$ are selected as PN sequence
and $\textit{\textbf{P}}_1=\textit{\textbf{P}}_2$. Assume that the
index of the training sequence is $l\in[0,2N_{P}-1]$, the frequency
domain training sequence can be expressed as
\begin{equation} \label{13}
D_{1,l}= \left\{ \begin{aligned}
         P_{1}(\lfloor l/2 \rfloor) &, ~l\in[0,2N_{P}-1] \\ 0~~~~~~~~~&, ~\textmd{else}
                          \end{aligned} \right.
                          \end{equation}
where $\lfloor \cdot \rfloor$ denotes the floor function.

\section{The Proposed Carrier Frequency Offset Estimation Algorithm}
We assume that the received signal ${r}(t)$ experiences a carrier
frequency offset $\Delta f$ and the timing offset is completely
compensated. The demodulated training sequences can be expressed as
\begin{equation} \label{14}
\begin{split}
\hat{D}_{1,2l}&=D_{1,2l}\Xi_{1,2l;1,2l}^{\Delta
 f}+\sum_{l^{'}=0,l^{'}\neq l}^{N/2-1}D_{1,2l^{'}}\Xi_{1,2l;1,2l^{'}}^{\Delta
 f}\\
&+\sum_{k^{'}\neq
1}\sum_{l^{'}=0}^{N/2-1}D_{k^{'},2l^{'}}\Xi_{k,2l;k^{'},2l^{'}}^{\Delta
 f}\\
&+\sum_{k^{'}}\sum_{l^{'}=0}^{N/2-1}D_{k^{'},2l^{'}+1}\Xi_{k,2l;k^{'},2l^{'}+1}^{\Delta
 f}+W_{0,2l}
\end{split}
\end{equation}
and
\begin{equation} \label{15}
\begin{split}
\hat{D}_{1,2l+1}&=D_{1,2l+1}\Xi_{1,2l+1;1,2l+1}^{\Delta
 f}\\
&+\sum_{l^{'}=0,l^{'} \neq
l}^{N/2-1}D_{1,2l^{'}+1}\Xi_{1,2l+1;1,2l^{'}+1}^{\Delta
 f}\\
&+\sum_{k^{'}\neq
1}\sum_{l^{'}=0}^{N/2-1}D_{k^{'},2l^{'}+1}\Xi_{k,2l+1;k^{'},2l^{'}+1}^{\Delta
 f}\\
&+\sum_{k^{'}}\sum_{l^{'}=0}^{N/2-1}D_{k^{'},2l^{'}}\Xi_{k,2l+1;k^{'},2l^{'}}^{\Delta
 f}+W_{0,2l+1}
\end{split}
\end{equation}
$W_{k,2l}$ and $W_{k,2l+1}$ denote the AWGN in frequency domain.
Thanks to the central limit theorem\cite{Pro07}, the last four terms
in (14) and (15) can be denoted as $\phi_{1,2l}$ and
$\phi_{1,2l+1}$, respectively. We can rewrite equation (14) and (15)
as
\begin{equation} \label{16}
\begin{split}
\hat{D}_{1,2l}&=D_{1,2l}\Xi_{1,2l;1,2l}^{\Delta f}+\phi_{1,2l}\\
&=D_{1,2l}e^{-j2\pi T'\Delta f}\int\int
h(\tau,\upsilon)A_{\psi}^{\ast}(\tau,\upsilon+\Delta f)\\
&\cdot e^{-j2\pi2lF'\tau}e^{j2\pi\upsilon T'}d\tau
d\upsilon+\phi_{1,2l}
\end{split}
\end{equation}
and
\begin{equation} \label{17}
\begin{split}
\hat{D}_{1,2l+1}&=D_{1,2l+1}\Xi_{1,2l+1;1,2l+1}^{\Delta f}+\phi_{1,2l+1}\\
&=D_{1,2l+1}e^{-j2\pi (1+\frac{1}{2})T'\Delta f}\int\int
h(\tau,\upsilon)\\
&\cdot A_{\psi}^{\ast}(\tau,\upsilon+\Delta
f)e^{-j2\pi(2l+1)F'\tau}\\
&\cdot e^{j2\pi\upsilon (1+1/2)T'}d\tau d\upsilon+\phi_{1,2l+1}
\end{split}
\end{equation}

We assume that $D_{1,2l}D_{1,2l}^{\ast}=\sigma_{s}^{2}$ and define
$\gamma_m$ as (18) at the top of next page. $\vartheta_0=\pi
T'\Delta f$ denotes the phase rotation caused by frequency offset
and $\vartheta_1$ is a constant. Let $\varepsilon=\Delta fT_sN/2$
denotes the normalized frequency offset, and $T_s$ denotes the
sample period. $W$ is the additive noise with zero mean and average
power $\sigma_{W}^2$. There are two fine frequency offset estimation
schemes: Phase differential estimation scheme and least squares
estimation scheme.

\begin{figure*}
\normalsize \setcounter{mytempeqncnt}{\value{equation}}
\setcounter{equation}{17}
\begin{equation} \label{18}
\begin{split}
\gamma_m&=\sum_{l=0}^{N/2-1}\hat{D}_{1,2l+2m+1}\textmd{exp}\big(j\cdot
\arg(D_{1,2l+2m+1}^{\ast})\big)\bigg(\hat{D}_{1,2l}\textmd{exp}\big(j\cdot
\arg(D_{1,2l}^{\ast})\big)\bigg)^{\ast}\\
&=\sum_{l=0}^{N/2-1}\bigg\{\sigma_{s}^{2}e^{-j\pi T'\Delta
f}\sum_{i=0}^{L-1}e^{-j2\pi
(2m+1)iF'}\int_{-f_d}^{f_d}S(i,\upsilon)\big|A_{\psi}^{\ast}(i,\upsilon+\Delta
f)\big|^{2}e^{j\pi\upsilon T'}d\upsilon +W_m\bigg\}\\
&\simeq
e^{-j\vartheta_0}e^{-j(m+1)\vartheta_1}\sum_{l=0}^{N/2-1}\bigg\{\sigma_{s}^{2}\bigg|\sum_{i=0}^{L-1}e^{-j2\pi
(2m+1)iF'}\int_{-f_d}^{f_d}S(i,\upsilon)\big|A_{\psi}^{\ast}(i,\upsilon+\Delta
f)\big|^{2}e^{j\pi\upsilon T'}d\upsilon \bigg| \bigg\}+W
\end{split}
\end{equation}
\setcounter{equation}{\value{mytempeqncnt}} \hrule
\end{figure*}

\subsection{Phase Differential Carrier Frequency Offset Estimation Algorithm}
The phase differential carrier frequency offset estimation algorithm
can be expressed as \setcounter{equation}{18}
\begin{equation} \label{19}
\begin{split}
\hat{\varepsilon}&=\frac{N}{2\pi
M}\bigg(\arg(\gamma_0)\\
&-\frac{1}{N/2-1}\sum_{m=1}^{N/2-1}\big(\arg(\gamma_{m})-\arg(\gamma_{m-1})
\big) \bigg)
\end{split}
\end{equation}
The phase differential estimation scheme is valid for
$\varepsilon\in[-N/2M,N/2M]=[-\rho/2,\rho/2]$, and $\rho$ denotes
system spectral efficiency.

\subsection{Least Squares Carrier Frequency Offset Estimation Algorithm}
We can model the phase of $\gamma_m$ as
\begin{equation} \label{20}
\begin{split}
\arg(\gamma_m)\simeq\vartheta_0+(m+1)\vartheta_1+\zeta_m
\end{split}
\end{equation}
and
\begin{equation} \label{21}
\begin{split}
\arg(\vec{\gamma})\simeq \mathbf{C} \vec{\vartheta}+\vec{\zeta}
\end{split}
\end{equation}
where $\vec{\gamma}=[\gamma_0,\gamma_1,\cdots,\gamma_{N/2-1}]$ and
$\arg(\vec{\gamma})=[\arg(\gamma_0),\arg(\gamma_1),\cdots,\arg(\gamma_{N/2-1})]^T$,
$\vec{\vartheta}=[\vartheta_0,\vartheta_1]$ and
$\vec{\zeta}=[\zeta_0,\zeta_1,\cdots,\zeta_{N/2-1}]$. $\mathbf{C}$
in (21) can be expressed as
\begin{equation} \label{22}
\mathbf{C} = \left[\begin{IEEEeqnarraybox*}[][c]{,c/c,}
1&0\\
1&1\\
\vdots&\vdots\\
1&N/2-1
\end{IEEEeqnarraybox*}\right]
\end{equation}
The least squares carrier frequency offset estimation algorithm can
be expressed as
\begin{equation} \label{23}
\begin{split}
\hat{\vec{\vartheta}}=(\mathbf{C}^T\mathbf{C})^{-1}\mathbf{C}^T\arg(\vec{\gamma})
\end{split}
\end{equation}
and $(\mathbf{C}^T\mathbf{C})^{-1}$ in (23) can be obtained by
\begin{equation} \label{24}
(\mathbf{C}^T\mathbf{C})^{-1}=\frac{2}{N\Delta m^2}
\left[\begin{IEEEeqnarraybox*}[][c]{,c/c,}
\overline{m^2}&-\overline{m}\\
-\overline{m}&1
\end{IEEEeqnarraybox*}\right]
\end{equation}
where $\overline{m}=2/N\sum_{k=0}^{N/2-1}k$,
$\overline{m^2}=2/N\sum_{k=0}^{N/2-1}k^2$ and $\Delta
m^2=2/N\sum_{k=0}^{N/2-1}(k-\overline{m})^2$. Hence, the estimation
of $\hat{\vec{\vartheta}}$ can be rewritten as
\begin{equation} \label{25}
\begin{split}
\hat{\vec{\vartheta}}=[\hat{\vartheta}_0,\hat{\vartheta}_1]^{T}=\frac{1}{\Delta
m^2}\left[\begin{IEEEeqnarraybox*}[][c]{,c/c,}
\overline{\gamma}\overline{m^2}&\overline{m}(\overline{\gamma m})\\
\overline{\gamma m}&\overline{m}\cdot\overline{\gamma}
\end{IEEEeqnarraybox*}\right]
\end{split}
\end{equation}
where $\overline{\gamma}=2/N\sum_{k=0}^{N/2-1}\arg(\gamma_k)$ and
$\overline{\gamma m}=2/N\sum_{k=0}^{N/2-1}\arg(\gamma_k k)$. The
phase rotation $\vartheta_0$ caused by frequency offset and the
constant $\vartheta_1$ can be obtained by
\begin{equation} \label{26}
\begin{split}
\hat{\vartheta}_1=\frac{\overline{\gamma
m}-\overline{m}\cdot\overline{\gamma}}{\Delta m^2}
\end{split}
\end{equation}
and
\begin{equation} \label{27}
\begin{split}
\hat{\vartheta}_0=\arg(\gamma_0)-\hat{\vartheta}_1=\frac{\overline{\gamma}\overline{m^2}-\overline{m}(\overline{\gamma
m})}{\Delta m^2}
\end{split}
\end{equation}

The normalized frequency offset $\hat{\varepsilon}$ can be expressed
as
\begin{equation} \label{28}
\begin{split}
\hat{\varepsilon}=\frac{N\big(\arg(\gamma_0)-\hat{\vartheta}_1\big)}{2\pi
M}
\end{split}
\end{equation}
The least squares frequency offset estimation scheme is valid for
$\varepsilon\in[-N/2M,N/2M]=[-\rho/2,\rho/2]$.

\subsection{Simplified Carrier Frequency Offset Estimation Scheme}
Let
${\mathbf{\Psi}}_{0}=[\psi_0(0),\psi_0(1),\cdots,\psi_0(L_\psi-1)]$
denotes the prototype pulse with length $L_\psi$ in discrete HMCT
system. In order to recover the transmitted frequency domain
training sequence $D_{1,l}$, we need $N_p$ inner product operations
with $L_\psi$ points, which introduce high computational complexity.

\newtheorem{theorem}{Theorem}
\begin{theorem}
Let $\textit{\textbf{r}}=[r(0),r(1),\cdots,r(L_{\psi}-1)]$ denote
the received signal vector with length $L_{\psi}$, and
$\mathbf{\Phi}=\{{\mathbf{\Psi}}_{n}\}$, $n=0,1,\cdots,N/2-1$,
denotes the objective projection subspace.
${\mathbf{\Psi}}_{n}=[\psi_n(0),\psi_n(1),\cdots,\psi_n(L_\psi-1)]$
and satisfies
\begin{equation} \label{29}
\begin{split}
\mathbf{\psi}_{n}(m)=\mathbf{\psi}_{0}(m)e^{\frac{j2\pi mn}{N/2}}
\end{split}
\end{equation}

The time frequency subspace projection of the received signal
$\textit{\textbf{r}}$ on the objective projection subspace
$\mathbf{\Phi}$ is equivalent to
$\textrm{FFT}\big(\sum_{l=0}^{Q_{\psi}-1}r_{\psi}(k+lN/2)\big)$,
$k=0,1,\cdots,N/2-1$. $Q_{\psi}=\lceil \frac{L_{\psi}}{N/2}\rceil$
and $\lceil \cdot \rceil$ denotes the ceiling function.
$\textit{\textbf{r}}_{\psi}=\textit{\textbf{r}}\odot\mathbf{\Psi}_0^{\ast}$
and
$\textit{\textbf{r}}_{\psi}=[r_{\psi}(0),r_{\psi}(1),\cdots,r_{\psi}(L_{\psi}-1)]$.
$\odot$ denotes the Hadamard product and $\textrm{FFT}(\cdot)$
denotes the Fast Fourier Transform.
\end{theorem}

\begin{IEEEproof}
The output symbol $Y(i)$ of the time frequency subspace projector
after projecting the received signal $\textit{\textbf{r}}$ to the
$i$th subspace $\mathbf{\Psi}_{i}$ can be expressed as
\begin{equation} \label{30}
\begin{split}
Y(i)&=\langle \textit{\textbf{r}},\mathbf{\Psi}_{i}\rangle=\sum_{n=0}^{L_{\psi}-1}r(n)\psi_{i}^{\ast}(n)\\
&=\sum_{n=0}^{L_{\psi}-1}r(n)\psi_{0}^{\ast}(n)e^{\frac{-j2\pi
ni}{N/2}}
\end{split}
\end{equation}
After performing the IFFT on the signal $Y(i)$, the transformed
signal can be expressed as
\begin{equation} \label{31}
\begin{split}
y(k)&=\sum_{i=0}^{N/2-1}Y(i)e^{\frac{j2\pi ki}{N/2}}\\
&=\frac{1}{N/2}\sum_{i=0}^{N/2-1}\sum_{n=0}^{L_{\psi}-1}r(n)\psi_{0}^{\ast}(n)e^{\frac{j2\pi
(k-n)i}{N/2}}\\
&=\frac{1}{N/2}\sum_{i=0}^{N/2-1}\sum_{n=0}^{N/2-1}\sum_{l=0}^{Q_{\psi}-1}r_{\psi}(n+\frac{lN}{2})e^{\frac{j2\pi
(k-n)i}{N/2}}\\
&=\sum_{l=0}^{Q_{\psi}-1}r_{\psi}(k+\frac{lN}{2})
\end{split}
\end{equation}
where $\lceil \cdot \rceil$ denotes the ceiling
 function. $r_{\psi}(n)=r(n)\psi_{0}^{\ast}(n)$,
 $n=0,1,\cdots,L_{\psi}-1$, which can be denoted as
 $\textit{\textbf{r}}_{\psi}=\textit{\textbf{r}}\odot\mathbf{\Psi}_0^{\ast}$.
\end{IEEEproof}

We can conclude from $\textit{Theorem 1}$ that the transmitted
frequency domain training sequence $D_{1,l}$ can be obtained by the
following two steps. The first step is to calculate the Hadamard
product of the received signal $\textit{\textbf{r}}$ and the
prototype pulse ${\mathbf{\Psi}}_{0}$, that is
$\textit{\textbf{r}}_{\psi}$. The second step is to superimpose
vector $\textit{\textbf{r}}_{\psi}$ with period $N/2$ and perform
$N/2$-point FFT on the superimposed vector
$\mathbf{y}=[y(0),y(1),\ldots,y(N/2)]$. Hence, two Hadamard product
operations, two superimpose operators with period $N/2$ and two
$N/2$-point FFT are needed to recover the transmitted frequency
domain training sequence $D_{1,l}$, which is a low complexity
approach compared to the traditional $N_p$ projectors approach.

\subsection{Cramer-Rao Lower Bound}
We can conclude from (19) and (28) that the proposed phase
differential and least squares carrier frequency offset estimators
are functions of $\gamma_0$. $\gamma_0$ in AWGN channel can be
expressed as
\begin{equation} \label{32}
\begin{split}
\gamma_0&=\sum_{l=0}^{N/2-1}\Bigg\{\sigma_{s}^{2}e^{-j\pi T'\Delta
f}+\phi_{1,2l+1}\sigma_se^{j2\pi T'\Delta
f}\\
&+\phi_{1,2l}^{\ast}\sigma_se^{-j2\pi(1+1/2)T'\Delta
f}+\phi_{1,2l+1}\phi_{1,2l}^{\ast} \Bigg\}\\
&=\frac{N}{2}\sigma_{s}^{2}e^{-j2\pi M\varepsilon/N}+W_0
\end{split}
\end{equation}
The conditional probability density function
$P(\gamma_0;\varepsilon)$ can be expressed as
\begin{equation} \label{33}
\begin{split}
P(\gamma_0;\varepsilon)&=\frac{1}{(\pi\sigma_{W}^2)^{1/2}}\textmd{exp}\Big\{
-\frac{1}{\sigma_{W_0}^2}\Big(\gamma_0-\frac{N}{2}\sigma_{s}^{2}e^{\frac{-j2\pi
M\varepsilon}{N}}\Big)\\
&\cdot\Big(\gamma_0-\frac{N}{2}\sigma_{s}^{2}e^{\frac{-j2\pi
M\varepsilon}{N}}\Big)^{\ast}\Big\}
\end{split}
\end{equation}
Differentiating the log likelihood function $\ln
P(\gamma_0;\varepsilon)$ with respect to $\varepsilon$, we have
\begin{equation} \label{34}
\begin{split}
\frac{\partial\ln
P(\gamma_0;\varepsilon)}{\partial\varepsilon}&=-\frac{1}{\sigma_{W_0}^2}\Big\{\Big(
\gamma_0-\frac{N}{2}\sigma_{s}^{2}e^{\frac{-j2\pi M\varepsilon}{N}}
\Big)\\
&\cdot\Big(-j\pi M\sigma_s^2e^{\frac{j2\pi
M\varepsilon}{N}}\Big)+\Big(
\gamma_0-\frac{N}{2}\sigma_{s}^{2}e^{\frac{-j2\pi M\varepsilon}{N}}
\Big)^{\ast}\\
&\cdot\Big(j\pi M\sigma_s^2e^{\frac{-j2\pi M\varepsilon}{N}}\Big)
\Big\}
\end{split}
\end{equation}
and
\begin{equation} \label{35}
\begin{split}
\frac{\partial^2\ln
P(\gamma_0;\varepsilon)}{\partial^2\varepsilon}&=-\frac{1}{\sigma_{W_0}^2}\Big(\frac{2\pi^2M^2\sigma_{s}^2}{N}\gamma_{0}e^{\frac{j2\pi
M\varepsilon}{N}}\\
&+\frac{2\pi^2M^2\sigma_{s}^2}{N}\gamma_{0}^{\ast}e^{\frac{-j2\pi
 M\varepsilon}{N}}\Big)
\end{split}
\end{equation}
Hence, the Cramer-Rao lower bound of the proposed carrier frequency
offset can be expressed as
\begin{equation} \label{37}
\begin{split}
\textmd{E}\{|\hat{\varepsilon}-\varepsilon|^2\}&\geq-E\Big[\frac{\partial^2\ln
P(\gamma_0;\varepsilon)}{\partial^2\varepsilon}
\Big]^{-1}\\
&=\frac{N}{2\pi^2M^2\textmd{SNR}}
\end{split}
\end{equation}

\section{Simulation Results}
In this section, we test the proposed synchronization approach for
MCM system with hexagonal T-F lattice via computer simulations based
on the discrete signal model. In the following simulations, the
number of subcarriers for HMCT system is chosen as $N$=40, and the
length of prototype pulse $L_{\psi}$=600. The center carrier
frequency is $f_c$=5GHz and the sampling interval is set to
$T_{s}$=$10^{-6}$s. The system parameters of HMCT system are
$F$=25kHz, $T$=$1\times10^{-4}$s and signaling efficiency is set to
$\rho$=0.8. WSSUS channel is chosen as DD channel with exponential
power delay profile and U-shape Doppler spectrum.

The MSE performance of the proposed carrier frequency estimation
algorithm over AWGN channel is given in Fig. 2. We can see from Fig.
2 that the proposed least squares carrier frequency estimation
algorithm outperforms phase differential scheme at low SNR, and both
the proposed schemes can obtain an approximation to the Cramer-Rao
lower bound MSE performance.

\begin{figure}[!t] \centering
\includegraphics[width=3in]{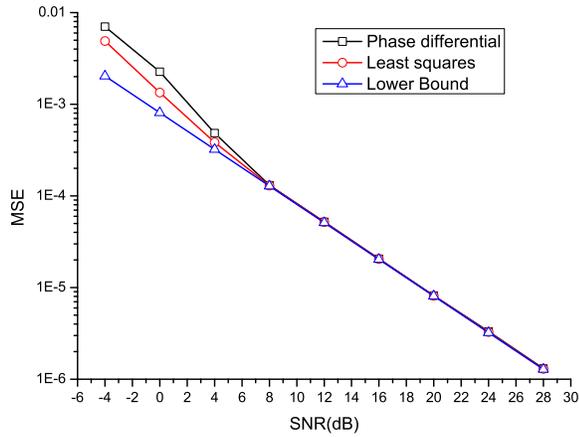}
\caption{MSE performance of the proposed carrier frequency
estimation algorithm over AWGN channel.}
\end{figure}

The MSE performance of the proposed carrier frequency estimation
algorithm over DD channel is given in Fig. 3. As shown in Fig. 3
that the proposed least squares carrier frequency estimation
algorithm outperforms phase differential scheme at low SNR, but the
phase differential scheme outperforms least squares algorithm at
high SNR. The DD propagation channel introduces energy perturbation
among the transmitted symbols, hence there is a gap between the MSE
performance of the proposed two carrier frequency offset estimation
algorithms and that of the Cramer-Rao lower bound.

\begin{figure}[!t] \centering
\includegraphics[width=3in]{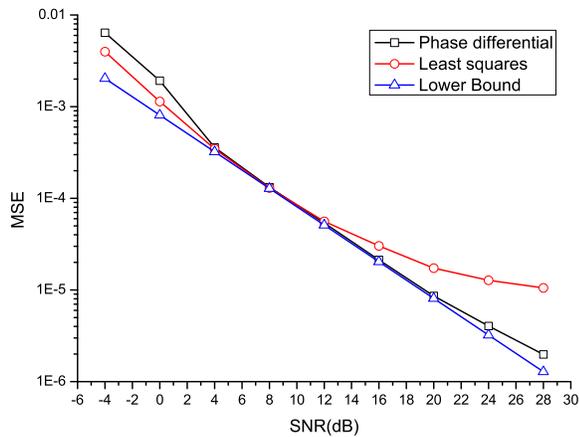}
\caption{MSE performance of the proposed carrier frequency
estimation algorithm over DD channel.}
\end{figure}

\section{Conclusion}
In this paper, the effects of carrier frequency offset on HMCT
system is analyzed. Meanwhile, a novel carrier frequency offset
estimation approach, including preamble structure, carrier frequency
offset estimation algorithm, is proposed for HMCT system. After
detailed derivation, the close form performance lower bound of the
proposed carrier frequency offset estimation alglrithm is given.
Theoretical analyses and simulation results show that the proposed
carrier frequency offset estimation approach for HMCT system obtains
an approximation to the Cramer-Rao lower bound MSE performance over
the DD channel with exponential power delay profile and U-shape
Doppler spectrum.

\section*{Acknowledgment}
This work was supported by the Jiangsu Province National Science
Foundation for young scholars under Grant (No. BK2012055) and the
Young Scientists Pre-research Fund of PLAUST under Grant (No.
KYTYZLXY1211), the National Natural Science Foundation of China (No.
60972050) and the Jiangsu Province National Science Foundation under
Grant (No. BK2011002).

\ifCLASSOPTIONcaptionsoff
  \newpage
\fi


\begin{thebibliography}{00}
\bibitem{Kumb07}
Kumbasar V., Kucur O. ICI reduction in OFDM systems by using
improved sinc power pulse. \emph{Digital Signal Processing}, vol.17,
no.6, pp.997-1006, Nov. 2007.
\bibitem{Das07}
Das S., Schniter P. Max-SINR ISI/ICI-Shaping multicarrier
communication over the doubly dispersive channel. \emph{IEEE
Transactions on Signal Processing}, vol.55, no.12, pp.5782-5795,
Dec. 2007.
\bibitem{Abb10}
Abbas H.K., Waleed A. M., Nihad S., The performance of multiwavelets
based OFDM system under different channel conditions \emph{Digital
Signal Processing}, vol.20, no.2, pp.472-482, Mar. 2010.
\bibitem{Gao11}
Gao X., Wang W., Xia X.G., et al. Cyclic prefixed OQAM-OFDM and its
application to single-carrier FDMA. \emph{IEEE Transactions on
Communications}, vol.59, no.5, pp.1467-1480, May 2011.
\bibitem{Jun07}
P. Jung, G. Wunder, The WSSUS pulse design problem in multicarrier
transmission, \emph{IEEE Transactions on Communications}, vol.55,
no. 10, pp.1918-1928, Oct. 2007.
\bibitem{Lin08}
G. Lin, L. Lundheim, N. Holte, Optimal pulses robust to carrier
frequency offset for OFDM/QAM systems, \emph{IEEE Communications
Letters}, vol. 12, no. 3, pp. 161-163, Mar. 2008.
\bibitem{Sio02}
P. Siohan, C. Siclet, N. Lacaille, Analysis and design of OFDM/OQAM
systems based on filterbank theory, \emph{IEEE Transactions on
Signal Processing}, vol. 50, no. 5, pp. 1170-1183, May 2002.
\bibitem{Ma08}
M. Ma, B. Jiao, C. Y. Lee, A dual-window technique for enhancing
robustness of OFDM against frequency offset, \emph{IEEE
Communications Letters}, vol. 12, no. 1, pp. 17-19, Jan. 2008.
\bibitem{Str03}
Strohmer T, Beaver S. Optimal OFDM design for time-frequency
dispersive channels. \emph{IEEE Trans. Commun.} vol.51, no.7,
pp.1111-1122, Jul. 2003.
\bibitem{Yua08}
Yuan Z.G., Shen Y.H., A novel LOFDM signal and its optimization over
doubly-dispersion channels. in \emph{Proc. 3rd IEEE Conference on
Industrial Electronics and Applications, 2008. ICIEA 2008},
pp.853-856, Jun. 2008.
\bibitem{Han07}
Han F.M., Zhang X.D. Hexagonal multicarrier modulation: A robust
transmission scheme for time-frequency dispersive channels.
\emph{IEEE Transactions on Signal Processing}, vol.55, no.5,
pp.1955-1961, May 2007.
\bibitem{Han09}
Han F.M., Zhang X.D., MLSD for hexagonal multicarrier transmission
with time-frequency localized pulses, \emph{IEEE Transactions on
Vehicular Technology}, vol.58, no.3, pp.1598-1604, Mar. 2009.
\bibitem{Han10}
Han F.M., Zhang X.D., Asymptotic Equivalence of Two Multicarrier
Transmission Schemes in Terms of Robustness Against Time-Frequency
Dispersive Channels. \emph{IEEE Transactions on Vehicular
Technology},vol.59, no.2, pp.1598-1604, Feb. 2010.
\bibitem{Xu09}
Xu K., Shen Y. H., Effects of carrier frequency offset, timing
offset, and channel spread factor on the performance of hexagonal
multicarrier modulation systems. \emph{EURASIP Journal on Wireless
Communications and Networking}, vol.2009, pp.1-8, Jan. 2009.
\bibitem{Xu11}
Xu K., Xu Y., Zhang D., SINR analysis of hexagonal multicarrier
transmission systems in the presence of insufficient synchronization
for doubly dispersive channel. \emph{Frequenz}, vol.65, no.5,
pp.149-157, Aug. 2011.
\bibitem{Xu12}
Xu K., Lv Z., Xu Y., Zhang D., Max-SINR Based Timing Synchronization
Scheme in Hexagonal Multicarrier Transmission. \emph{Wireless
Personal Communications}, DOI: 10.1007/s11277-012-0550-5, 2012.
\bibitem{Xu12-1}
Xu K., Xu Y., Xia X., Zhang D., On Max-SINR Receiver for Hexagonal
Multicarrier Transmission Over Doubly Dispersive Channel. to appear
in \emph{Proc. IEEE GLOBECOM 2012}.
\bibitem{Bel63}
Bello P. A.,Characterization of randomly time-variant linear
channels, \emph{IEEE Transactions on Communication System},
 vol.11, no.4, pp.360-393, Dec. 1963.
\bibitem{Coh95}
L. Cohen, \emph{Time-frequency analysis}. Englewood Cliffs, NJ:
Prentice-Hall, 1995.
\bibitem{Mat02}
P. Matthias, \emph{Mobile fading channels}. West Sussex, England:
John Wiley \& Sons, Ltd, 2002.
\bibitem{Con98}
J. H. Conway and N. J. A. Sloane, \emph{Sphere Packings, Lattices
and Groups}, 3rd ed. New York: Springer-Verlag, 1998.
\bibitem{Pro07}
J. Proakis and M. Salehi, \emph{Digital Communications}, 5th
edition. New York: McGraw-Hill, 2007.

\end{thebibliography}
\end{document}